\documentclass[aps,letterpaper,twocolumn,pra,footinbib,floatfix,showpacs]{revtex4}
\usepackage{amsmath}
\usepackage{amsfonts}
\usepackage{amssymb}
\usepackage[scanall]{psfrag}
\usepackage{psfrag}
\usepackage{graphicx}
\usepackage{color}

\begin{document}
\newcommand{\of}[1]{\left( #1 \right)}
\newcommand{\sqof}[1]{\left[ #1 \right]}
\newcommand{\abs}[1]{\left| #1 \right|}
\newcommand{\avg}[1]{\left< #1 \right>}
\newcommand{\cuof}[1]{\left \{ #1 \right \} }
\newcommand{\bra}[1]{\left < #1 \right | }
\newcommand{\ket}[1]{\left | #1 \right > }
\newcommand{\pil}{\frac{\pi}{L}}
\newcommand{\bx}{\mathbf{x}}
\newcommand{\by}{\mathbf{y}}
\newcommand{\bk}{\mathbf{k}}
\newcommand{\bp}{\mathbf{p}}
\newcommand{\bl}{\mathbf{l}}
\newcommand{\bq}{\mathbf{q}}
\newcommand{\bs}{\mathbf{s}}
\newcommand{\psibar}{\overline{\psi}}
\newcommand{\svec}{\overrightarrow{\sigma}}
\newcommand{\dvec}{\overrightarrow{\partial}}
\newcommand{\bA}{\mathbf{A}}
\newcommand{\bdelta}{\mathbf{\delta}}
\newcommand{\bK}{\mathbf{K}}
\newcommand{\bQ}{\mathbf{Q}}
\newcommand{\bG}{\mathbf{G}}
\newcommand{\bw}{\mathbf{w}}
\newcommand{\bL}{\mathbf{L}}
\newcommand{\ohat}{\widehat{O}}
\newcommand{\up}{\uparrow}
\newcommand{\down}{\downarrow}
\newcommand{\MM}{\mathcal{M}}

\author{Eliot Kapit}
\affiliation{Initiative for Theoretical Science, The Graduate Center, City University of New York, 365 5th Avenue, New York, NY 10016}

\title{Universal two-qubit interactions, measurement and cooling for quantum simulation and computing}

\begin{abstract}

By coupling pairs of superconducting qubits through a small Josephson junction with a time-dependent flux bias, we show that arbitrary interactions involving any combination of Pauli matrices can be generated with a small number of drive tones applied through the flux bias of the coupling junction. We then demonstrate that similar (though not fully universal) results can be achieved in capacitively coupled qubits by exploiting the higher energy states of the devices through multi-photon drive signals applied to the qubits' flux degrees of freedom. By using this mechanism to couple a qubit to a detuned resonator, the qubit's rotating frame state can be non-destructively measured along any direction on the Bloch sphere. Finally, we describe how the frequency-converting nature of the couplings can be used to engineer a mechanism analogous to dynamic nuclear polarization in NMR systems, capable of cooling an array of qubits well below the ambient temperature, and outline how higher order interactions, such as local 3-body terms, can be engineered through the same couplings. Our results demonstrate that a programmable quantum simulator for large classes of interacting spin models could be engineered with the same physical hardware.

\end{abstract}

\maketitle

\section{Introduction}

Quantum computing promises revolutionary improvements over traditional classical algorithms to solve hard problems, provided that a functional, noise-tolerant quantum computer can be built. A number of paradigms for engineering a quantum computer have been considered in recent years, with the most studied being the ``gate model," where the Hamiltonian of a system of $N$ quantum bits (two-level systems) is assumed to be trivial and all unitary evolution of the system wavefunction occurs through a digital series of finite-length pulses to enact specific unitary operators. A key fact which makes such constructions feasible is that the sequential combination of single-qubit rotations with a two-qubit interaction suffices to generate all possible two-qubit unitary transformations \cite{divincenzobacon2000}. This allows arbitrary quantum operations to be performed through a small set of distinct physical terms, greatly simplifying the underlying system architecture.

However, in many cases it is desirable (or unavoidable) for the system to have a nontrivial physical Hamiltonian $H$ that contributes continuously to its time evolution, and since single-qubit rotations generically do not commute with $H$, performing arbitrary deformations of the system wavefunction through a small handful of discrete gates becomes difficult or impossible. Further, in the adiabatic model of quantum computing \cite{farhigoldstone2001}, only slow variations of the system Hamiltonian are allowed and thus all unitary time evolution must occur from the application of a continuous, physical $H \of{t}$. Quantum computing with physical Hamiltonians has a number of potential advantages over the pure gate model, including simplicity of the algorithms \cite{farhigoldstone2000}, resilience to low frequency noise and small variations in the system parameters (often through topological protection \cite{nayaksimon,fowlersurface}), and even passive correction of quantum errors arising from unwanted interactions with the environment \cite{youngsarovar2013,sarovaryoung2013,kapitchalker2014,brownloss2014}. Further, such Hamiltonians could be tuned to mimic the structure of interesting spin models in condensed matter physics, directly simulating exotic condensed matter systems in a similar fashion to cold atoms in optical lattices \cite{bloch}. A simple method for continuously generating arbitrary multi-qubit Hamiltonians is thus extremely desirable.

In this article, we demonstrate that in circuit QED systems of superconducting qubits, a Josephson junction coupling with a time-dependent flux bias serves as a universal interaction resource, and can generate arbitrary qubit Hamiltonians of the form $\sum_{a,b}^{\cuof{x,y,z}} c_{ab} \sigma_{1}^{a} \sigma_{2}^{b}$, where the nine $c_{ab}$ coefficients generate all possible two-body interactions. We then show that by capacitively coupling a qubit to a resonator and applying appropriate multi-photon signals through the qubit's flux line, we can non-destructively measure the rotating frame state of the qubit along an arbitrary direction on the Bloch sphere. This technique does not require that the many-body Hamiltonian is turned off during the measurement, allowing aspects of the system to be probed without disrupting the many-body state. We also propose a simple mechanism based on these couplings which can cool a grid of qubits far below the ambient temperature of their environment. Appropriately tuned, this scheme could be used to aid the operation of an adiabatic quantum computer. Finally, we demonstrate that frequency-converting couplings (which promote a tunneling photon up or down in energy as it hops between qubits) can be used to effectively eliminate the qubit nonlinearities, generating higher-order interactions than simple 2-body terms.

\section{Universal Two-Qubit Hamiltonians}\label{universalsec}

\begin{figure}
\includegraphics[width=3.25in]{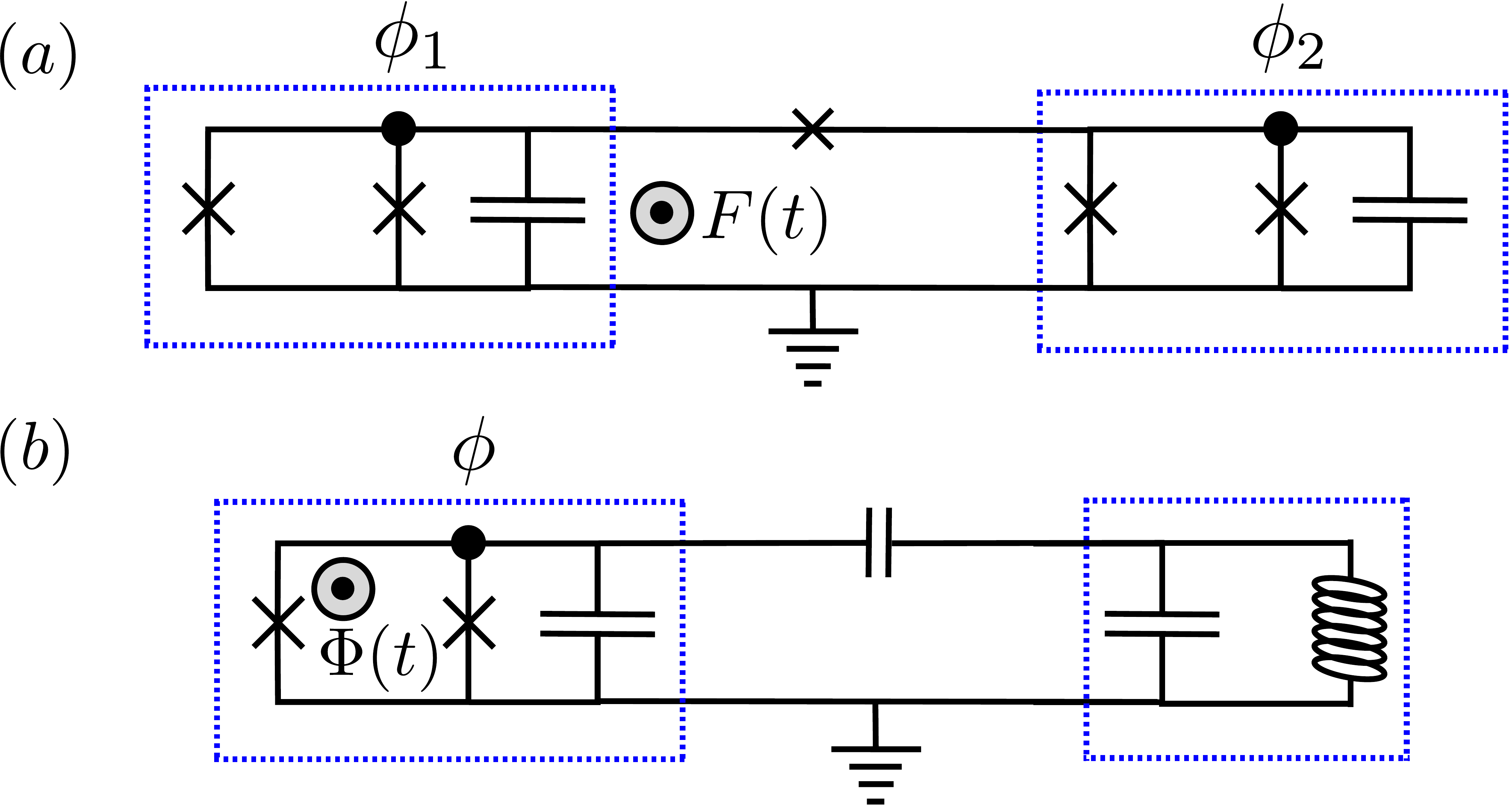}
\caption{Basic coupling types considered in this work. In figure (a), we demonstrate a flux biased Josephson junction coupling between two transmon qubits (blue dashed regions), discussed in Sec.~\ref{universalsec}, each of which has its primary Josephson junction split into a SQUID loop for optional flux tuning. By tuning the flux signal $F \of{t}$ threaded through the loop made by the smaller junction and the bridged ground, all possible two-body interactions between the qubits can be dynamically generated. Similar results can be derived for flux and fluxonium qubits in the appropriate parameter regimes. This coupling can be used to implement the cooling protocol of Sec.~\ref{coolsec}. In (b) we depict a transmon qubit coupled to a resonator through a simple capacitive coupling, as discussed in Sec.~\ref{capsec}. The grounds are bridged in this figure, though unlike case (a) this is not required to obtain the desired behavior. A flux signal is applied through the transmon's SQUID loop to generate couplings which allow for the transmon's rotating frame state to be measured along any direction on the Bloch sphere. This setup can also be used to couple qubits, though without as much flexibility as the small junction in (a).}\label{main fig}
\end{figure}

We consider a pair of charge insensitive superconducting qubits, such as transmon, flux or fluxonium devices \cite{Schoelkopf:2008p8712,Stajic:2013dh}. These qubits have one or more quantum phase degrees of freedom $\phi$, and when operated at a flux symmetry point and restricted to the lowest two levels, the phase operators take the generic form
\begin{eqnarray}\label{opids}
\sin \phi \to s \of{\sigma^+ + \sigma^-}, \; \cos \phi \to c_0 + c \sigma^z. 
\end{eqnarray}
The charge operator $Q$ becomes $\sigma^y$ and $z$ is the natural rest frame axis of the qubit. For devices such as flux qubits with two or more degrees of freedom, the phase operators can still nearly always be written in this form, with additional coefficients and sign flips if necessary. We now couple two qubits together, wiring them such that they share the same physical ground and are otherwise coupled by a Josephson junction of strength $\alpha E_J$ (where $E_J$ is the internal Josephson energy of the qubits), as first studied in \cite{gellerdonate2014,barendskelly2014}, with a time dependent flux bias $F \of{t}$. The grounds of the two circuits are bridged so that charge may flow through the devices and coupling junction in a closed loop, as in FIG. 1. We assume that the flux is solely threaded through the coupling region and does not leak into any internal loops in the qubit devices, and finally assume that the inductances created by bridging the grounds are negligible. Appealing to the circuit quantization equations for superconducting devices \cite{devoret1995}, the interaction term takes the form
\begin{eqnarray}\label{bareint}
H_{int} = -\alpha E_J \cos \of{\phi_1 - \phi_2 - F \of{t}}.
\end{eqnarray}

We assume that $\alpha$ is small so that in the absence of any applied drive tones the qubits do not interact and the distortion in the qubits' energy levels is minimal \cite{gellerdonate2014}. The bare excitation energies of the qubits are $\omega_1$ and $\omega_2$. Anticipating that we will drive the system through $F \of{t}$, we will work in a resonantly rotating frame through the transformation $\ket{\Psi} \to \exp \sqof{i \of{ \frac{\omega_1}{2} \sigma_{1}^{z} + \frac{\omega_2}{2} \sigma_{2}^{z} } t  }  \ket{\Psi}$. This in turn adds time dependence to the qubit operators:
\begin{eqnarray}
\sigma_{j}^{z} \to \sigma_{j}^{z}, \; \sigma_{j}^{\pm} \to \sigma_{j}^{\pm} e^{\mp i \omega_j t}.
\end{eqnarray}
We now choose:
\begin{eqnarray}\label{signal}
F \of{t} &=& \pi/2 - f_{zz} - 2  \left (f_{xy}^{(2)} \cos \sqof{ \of{\omega_1 + \omega_2 } t + \chi_1 + \chi_2 } \right. \nonumber \\
& & \left. + f_{xy}^{(0)}  \cos \sqof{ \of{\omega_1 - \omega_2 } t + \chi_1 - \chi_2 } \right )  \\
& & + 2 \sqrt{2} \of{ f_{xz} \cos \frac{\omega_1 t + \psi_1}{2}     + f_{zx} \cos  \frac{\omega_2 t + \psi_2}{2}    } . \nonumber
\end{eqnarray}
Here, the $\chi$ and $\psi$ terms are constant phase shifts, the coefficient $\beta$ controls the relative magnitude of the 2-photon and frequency-converting terms, and we assume all four of $f_{zz}, f_{xy}, f_{xz}$ and $f_{zx}$ are small compared to unity. Expanding the interaction term, we have
\begin{eqnarray}
H_{int} &=& - \alpha E_J \left[ \of{ \cos \phi_1 \cos \phi_2  + \sin \phi_1 \sin \phi_2 } \cos F \of{t} \right. \nonumber\\
& & \left. + \of{ \cos \phi_2 \sin \phi_1 - \cos \phi_1 \sin \phi_2  } \sin F \of{t}         \right]. 
\end{eqnarray}
We will now expand the terms in $H_{int}$ to lowest nontrivial order in the $f$'s, keeping only the terms which are independent of time and neglecting everything which is rapidly oscillating. The result for $f_{xy}^{(0)} = f_{xy}^{(2)}$ is:

\begin{widetext}
\begin{center}
\begin{table}
\begin{tabular}{| c || c | c |}
\hline
Signal & Rotating frame coupling & Role \\
\hline
$f_{zz}$ (Constant bias) & $\of{c_0 + c \sigma_{1}^{z} } \of{c_0 + c \sigma_{2}^{z}}$ & $zz$ interaction \\
$ \cos \sqof{ \of{\omega_1 - \omega_2 } t + \chi_1 - \chi_2 } $  & $e^{-i \of{\chi_{1}-\chi_{2}}} \sigma_{1}^{+} \sigma_{2}^{-} +  e^{i \of{\chi_{1}-\chi_{2}}} \sigma_{1}^{-} \sigma_{2}^{+}$ & Complex photon hopping/frequency conversion \cite{zakkabajjani2011} \\
$ \cos \sqof{ \of{\omega_1 + \omega_2 } t + \chi_1+ \chi_2 } $  & $e^{-i \of{\chi_{1}+\chi_{2}}} \sigma_{1}^{+} \sigma_{2}^{+} +  e^{i \of{\chi_{1}+\chi_{2}}} \sigma_{1}^{-} \sigma_{2}^{-}$ & 2-photon coherent pump \\
$ f_{xz} \cos \frac{\omega_1 t + \psi_1}{2} $ & $ \of{e^{- i \psi_{1}} \sigma_{1}^{+} + e^{+ i \psi_{1}} \sigma_{1}^{-}} \of{c_0 + c \sigma_{2}^{z}} $ & State-sensitive pump (continuous CNOT) \\ 
\hline
\end{tabular}
\caption{Fundamental interactions generated from driving a flux-biased Josephson junction with the signal (\ref{signal}). As shown in the text and equation (\ref{JJcoupling}), the nine free parameters of the signal allow for all two-body Hamiltonians $\sum_{a,b}^{\cuof{x,y,z}} c_{ab} \sigma_{1}^{a} \sigma_{2}^{b}$ to be generated from the same fixed device layout, FIG.~1a. Single qubit $z$ terms can be eliminated by detuning the drive signal(s) or applying flux biases to the qubits' internal flux lines; $x$ and $y$ terms can be eliminated by driving the qubits individually. }\label{inttab}
\end{table}
\end{center}

\begin{eqnarray}\label{JJcoupling}
H_{int} &=& - \alpha E_J \left[ f_{zz} \of{ c_0 + c \sigma_{1}^{z} } \of{c_0+ c \sigma_{2}^{z}} + f_{xy}^{(0)} \of{ \cos \chi_1 \sigma_{1}^{x} + \sin \chi_1 \sigma_{1}^{y} } \of{ \cos \chi_2 \sigma_{2}^{x} + \sin \chi_2 \sigma_{2}^{y} } \right. \\ 
& & \left. + f_{xz}^{2} \of{\cos \psi_{1} \sigma_{1}^{x} + \sin \psi_{1} \sigma_{1}^{y} } \of{ c_0 + c \sigma_{2}^{z} }   + f_{zx}^{2} \of{\cos \psi_{2} \sigma_{2}^{x} + \sin \psi_{2} \sigma_{2}^{y} } \of{ c_0 + c \sigma_{1}^{z} }     \right] + O \of{f^3}. \nonumber \\
& = & \sum_{a,b}^{\cuof{x,y,z}} c_{ab} \sigma_{1}^{a} \sigma_{2}^{b} + (\rm{single \; qubit \; terms}). \nonumber
\end{eqnarray}
\end{widetext}

The single qubit terms can be cancelled out by combinations of detunings and single photon drive fields, if desired. Counting degrees of freedom (including the relative magnitude of the 2-photon and frequency-converting parts of the $f_{xy}$ signal, which we set equal to 1 to make the above expression more compact), we see that we have nine independently adjustable parameters, sufficient to independently tune all nine of the $c_{ab}$. We note that in the more general case where the qubits have internal flux biases which are not $0$ or $\pi$ (and time reversal symmetry is thus broken), instead of using the trigonometric operator identifications (\ref{opids}) it can be more convenient to expand the coupling in the qubit phases by letting $\phi = \phi_0 + \alpha \of{\sigma^+ + \sigma^-}$, where $\phi_0$ is the minimum of the biased Josephson potential in each qubit. One will arrive at the same result in either case.

We have thus demonstrated that a flux biased Josephson junction is a truly universal interaction resource for quantum Hamiltonian engineering in superconducting devices. A wide array of interesting spin models can be implemented through these couplings-- for example, Kitaev's honeycomb \cite{kitaev} model can easily be engineered in both the abelian and non-abelian parameter regimes, though the question of how to prepare and maintain the ground state in either case is subtle and difficult. We note also that neglecting rapidly oscillating terms requires that all matrix elements are small compared to $\omega_1/2$, $\omega_2/2$, $\abs{\omega_1 - \omega_2 }$, $\omega_1 + \omega_2$ and so forth. Consequently, the qubit detunings and nonlinearities set the ultimate limit on the strength of interactions which can be generated through this method. A second limit is that continuous application of microwave fields will slowly heat the device by creating quasiparticle excitations; however, as the Cooper pair breaking energy in commonly used superconductors such as aluminum or titanium nitride is more than an order of magnitude larger than typical drive frequencies in this setup, quasiparticle generation will be suppressed. 

As a final note, as first pointed out in \cite{kapitgauge,hafeziadhikari}, the frequency-converting couplings \cite{zakkabajjani2011} (which conserve total $z$ spin and tunnel a photon from a qubit to one of its neighbors) can be used to engineer an artificial gauge field, where the electrically neutral photon excitations of the qubits behave as if they were charged particles moving in a magnetic field. Couplings produced by drive signals with phase delays (the $\chi_1$ and $\chi_2$ phases in Eq.~\ref{JJcoupling}) become complex in the rotating frame, and if a closed loop is formed by couplings where the phases $\chi$ do not sum to an integer multiple of $2 \pi$, then an artificial gauge field exists and the loop is pierced by an effective magnetic flux. This can be used to engineer single photon circulators \cite{kochhouck}, and in large lattices, fractional quantum Hall states of light \cite{kapithafezi2014}.

\section{Alternative formulation for capacitively coupled qubits: arbitrary non-destructive measurements in the rotating frame}\label{capsec}

The ability to engineer arbitrary interactions in the rotating frame is of limited usefulness unless it is coupled with an equivalent capacity for making arbitrary measurements of the superconducting qubits. In few-qubit experiments (or in purely gate-based quantum computing), arbitrary measurements can always be performed by simply turning off all the couplings at a chosen point in time and then using single-qubit rotations to rotate the qubits onto a natural measurement axis (conventionally, $\sigma^z$) before measuring them through standard means. However, in larger systems, the requirement of turning off the qubit Hamiltonian is a severe constraint, and is particularly destructive to continuous error correction schemes (which are generally sensitive to the energy levels of the many-body Hamiltonian in such systems). We therefore present a simple mechanism for non-destructively measuring the rotating frame state of single qubits, using a capacitive coupling to a resonator. Of course, one can achieve the same thing through a flux biased Josephson junction; however, capacitive couplings are substantially simpler to engineer (and do not require the grounds to be bridged). We will do so by exploiting the higher excited states of the transmon qubit; see for example \cite{chowgambetta2013,chenneill2014,solenoveconomou2014} for previous techniques which exploit this structure.

As before, we consider a superconducting qubit $T$, which we will treat as a three level system with nonlinearity $-\delta$. For transmon qubits, $\delta$ is positive and roughly equal to the charging energy $E_C$; for flux and fluxonium qubits, $\delta$ is negative and substantially larger. The qubit is capacitively coupled to a resonator $R$, and our goal is to engineer the rotating frame Hamiltonian $H' = \Lambda \mathbf{h} \cdot \sigma_T \of{a_{R}^{\dagger} + a_R}$. By measuring the voltage signal at the resonator, we can thus measure the dressed state of the qubit. We let the strength of the capacitive coupling between the qubit and resonator be $g$, and let the detuning $\omega_R - \omega_T \equiv \Delta$ be large. The system's Hamiltonian is
\begin{eqnarray}
H &=& \omega_T a_{T}^\dagger a_T + \omega_R a_R^\dagger a_R - \frac{\delta}{2} a_T^\dagger a_T^\dagger a_T a_T \\ & & + g \of{ a_T^\dagger a_R + a_R^\dagger a_T }. \nonumber
\end{eqnarray}
Before applying any drive fields, we want to diagonalize this Hamiltonian, for up to two photons in the transmon and $\cuof{n,n+1,n-1}$ in the resonator, to lowest order in $g/\Delta$. In the occupation basis $\ket{n_T, n_R}$, to lowest order in $g/\Delta$ we obtain
\begin{eqnarray}
\ket{\underline{0,n}} &=& \ket{0,n} - \frac{g}{\Delta} \ket{1,n-1} \sqrt{n}; \quad \epsilon \simeq n \omega_R \\
\ket{\underline{1,n}} &=& \ket{1,n} - \frac{q_2 g}{\Delta + \delta} \ket{2,n-1} \sqrt{n} \nonumber \\ & &+ \frac{g}{\Delta} \ket{0,n+1} \sqrt{n+1}; \quad \epsilon \simeq n \omega_R + \omega_T. \nonumber
\end{eqnarray}
Here, $q_2$ is the bose enhancement of the matrix element to add a second photon to the qubit ($q_2 \to \sqrt{2}$ in the limit of zero nonlinearity). We ignore $\ket{\underline{2,n}}$ as we are going to perturbatively eliminate the second level to reduce to the spin basis. We now apply the following combination of drive fields through the transmon's flux line (quantum phase variable $\phi$):
\begin{eqnarray}\label{Hdrive}
H_d &=& 2 \cos \phi \of{ f_1 \cos \sqof{ \of{\omega_R + \omega_T} t + \chi } + f_2 \cos \sqof{\Delta t - \chi} } \nonumber \\ & &+ 2 f_3 \sin \phi \cos \sqof{ \omega_R t }.
\end{eqnarray}
We now make the operator identifications:
\begin{eqnarray}\label{ident}
\cos \phi & \to & c_0 + c a_T^\dagger a_T + c_2 \of{ \ket{n+2,m}\bra{n,m} + {\rm H.c.} } \\
\sin \phi & \to & s_1 \of{ \ket{1,m}\bra{0,m} + {\rm H.c.} } + s_2 \of{\ket{2,m} \bra{1,m} + {\rm H.c.} }  \nonumber
\end{eqnarray}
The explicit flux signal required to generate $H_d$ depends on the details of the qubit. For example, in a split transmon \cite{kochyu} with parallel Josephson junctions $E_{J1}$ and $E_{J2}$, and no static flux bias, starting from $H_d = - E_{J1} \cos \of{\phi - \Phi \of{t}/2} - E_{J2} \cos \of{ \phi + \Phi \of{t}/2 }$ we get (all three of $k_1,k_2,k_3$ are assumed to be small):
\begin{widetext}
\begin{eqnarray}\label{Hphi}
\Phi \of{t} &=& 2 \sqof{ k_1 \cos \frac{ \of{\omega_R + \omega_T} t + \chi  }{2} + k_2 \cos \frac{ \of{\omega_R - \omega_T} t - \chi  }{2} + k_3 \cos \omega_R t  }; \\
H_d & = & - \of{E_{J1} + E_{J2}} \cos \phi \nonumber + \of{E_{J1} - E_{J2}} k_3 \sin \phi \cos \omega_R t + \of{\mathrm{off \; resonant}}  \\
& & + \of{E_{J1} + E_{J2}} \cos \phi \sqof{ \frac{k_{1}^{2}+k_{2}^{2}+k_{3}^{2}}{2} - \frac{k_{1}^{2}}{2} \cos \of{ \of{\omega_R + \omega_T} t + \chi } - \frac{k_{2}^{2}}{2} \cos \of{ \of{\omega_R - \omega_T} t - \chi }   } + O \of{k^{4}}. \nonumber
\end{eqnarray}
\end{widetext}
One can then read off the coefficients $f_1, f_2, f_3$ in (\ref{Hdrive}) from (\ref{Hphi}). We also obtain a term proportional of the form $k_3^{2} \cos \phi \cos 2 \omega_R t$; however, this term will only participate at higher orders in perturbation theory and will have a negligible impact.

We now move to the resonantly rotating frame where the qubit rotates at $\omega_T$ and the resonator at $\omega_R$. We first note that
\begin{eqnarray}
\bra{\underline{1,n-1}} \cos \phi \ket{\underline{0,n}} &=& -\frac{ c_0 g \sqrt{n}}{\Delta}, \nonumber \\
\bra{\underline{1,n+1}} \cos \phi \ket{\underline{0,n}} &=& - \frac{q_2 g c_2 \sqrt{n+1}}{\Delta+\delta}, \nonumber \\
\bra{\underline{0,n+1}} \sin \phi \ket{\underline{0,n}} &=& - \frac{s_{1} g \sqrt{n+1}}{\Delta}, \nonumber \\
\bra{\underline{1,n+1}} \sin \phi \ket{\underline{1,n}} &=& - \frac{q_2 s_{2} g \sqrt{n+1}}{\Delta + \delta} + \frac{s_1 g \sqrt{n+1}}{\Delta}. \nonumber 
\end{eqnarray}
Assuming all the frequencies/detunings are large compared to the Rabi frequencies of the drive fields, plugging these values into $H_d$ and discarding all time-dependent terms, we have:
\begin{eqnarray}
H' & = & - \frac{c_0 f_2 g  }{\Delta} \of{a_{R}^\dagger \sigma^- e^{i \chi} + \sigma^+ a_R e^{-i \chi}} \\ & & - \frac{q_2 g c_2 f_2}{\Delta + \delta } \of{a_{R}^\dagger \sigma^+ e^{-i \chi} + \sigma^- a_R e^{i \chi}} \nonumber \\
& & -\frac{g f_3}{2} \sqof{ \frac{q_2 s_2 }{\Delta + \delta}  + \of{ \frac{q_2 s_2 }{\Delta + \delta} - \frac{2 s_1}{\Delta} } \sigma^z } \of{a_{R}^\dagger + a_R}. \nonumber
\end{eqnarray}
Aside from a spin-independent $\of{a_{R}^\dagger + a_R}$ term which can be canceled out by an oscillating voltage $V Q_R \sin \omega_R t $ applied to the resonator, this is precisely $H' = \Lambda \of{ \mathbf{h} \cdot \sigma_T } \of{a_{R}^{\dagger} + a_R}$ as desired. This mechanism can thus be used to measure the $x$ and $y$ components of the qubits' rotating frame states, allowing for dressed state measurement without having to turn off the many-qubit Hamiltonian. 

Note also that this formulation can be applied to couple pairs of qubits, and generate the $x$, $y$, $zx$ and $zy$ components of $\sum_{a,b}^{\cuof{x,y,z}} c_{ab} \sigma_{1}^{a} \sigma_{2}^{b}$. The $zz$ components from this method, however, will be very small and are not easily tuned (they arise from the dispersive energy shift of each qubit based on the state of the other, and thus can only be tuned through adjusting the relative qubit detuning), so it is less versatile than the flux biased JJ of the previous section. A further complication arises when one wishes to engineer couplings of different types between a single qubit and many others; for the flux biased JJ, the fact that the coupling is specified entirely by the flux through the loop between two qubits means that couplings between each of the two qubits and anything else are irrelevant to that term in the Hamiltonian. In contrast, for capacitively coupled qubits, the coupling is generated by driving an internal degree of freedom in the qubit itself, which generically will influence the couplings between that qubit and any other degrees of freedom it interacts with. Finally, we note that an additional qubit $A$ can also be used for measurement through this mechanism; if the pulse $H' = \Lambda \of{ \mathbf{h} \cdot \sigma_T } \sigma_{A}^{x}$ is applied for a known interval then the operator $\mathbf{h} \cdot \sigma_T$ can be measured simply by measuring $\sigma_{A}^{z}$.

Having outlined two routes to universal interactions between pairs of superconducting qubits, we will now discuss two applications: passive cooling through frequency conversion, effectively replicating the NMR process of dynamic nuclear polarization in superconducting qubits, and higher-order interactions, where a single two-level qubit can be mapped to a $n$-level device (where the first $n$ levels are linearly spaced and there is a large nonlinearity to level $n+1$). Higher-order interactions can be used to engineer complex anyon models \cite{nayaksimon,kapitsimon,kapithafezi2014} with non-abelian exchange statistics, so they are of considerable interest to both condensed matter and quantum information physics.

\section{Dynamic Nuclear Polarization in Qubit Arrays}\label{coolsec}

As a simple application of the results of this work, we can use frequency-converting couplings as a dynamical cooling mechanism for groups of superconducting qubits. Specifically, we will demonstrate that by coupling a group of qubits to a second array of lossy qubits with significantly higher energies, we can engineer a passive cooling mechanism analogous to dynamic nuclear polarization in NMR \cite{malydebelouchina2014}, where magnetic polarization can be transferred from electrons to nuclei, whose vastly smaller gyromagnetic moment makes them otherwise difficult to polarize at achievable temperatures. Our mechanism could also be used, with some restrictions, to cool many-body systems below ambient temperature (as was done in an earlier experiment using a protocol which was specific to flux qubits  \cite{valenzuelaoliver2006}), and thus could be useful for quantum simulation and adiabatic quantum computing.

For simplicity, we consider identical and uncoupled primary qubits $P$, with system Hamiltonian $H = \omega \sum_j  \sigma_j^z /2$. In a thermal environment, the average spin $\avg{\sigma^z} = - \tanh \of{\omega / k_B T}$, and we wish to cool the system to a lower effective temperature without actively measuring the system, increasing $\omega$ or decreasing the environment temperature $T$. To do so, we introduce a second, ``shadow" lattice of auxiliary qubits $S$ \cite{kapithafezi2014,kapitchalker2014}, which have excitation energies $\omega_S$ and a fast relaxation rate $\Gamma_S$. To couple the two lattices, we introduce a frequency converting coupling as in (\ref{JJcoupling}), giving the total system Hamiltonian
\begin{eqnarray}\label{Hfreq}
H &=& - \sum_{j} \sqof{ \frac{\omega}{2} \sigma_j^z + \frac{\omega_S}{2} \sigma_{jS}^z } \\
& & + 2 g \sum_j \of{\sigma_{j}^+ \sigma_{jS}^- + \sigma_{j}^- \sigma_{jS}^+} \cos \Delta t, \nonumber
\end{eqnarray}
where $\Delta = \omega_S - \omega$ and we assume $g \ll \omega, \Delta$. As before, we work in a resonantly rotating frame by applying the unitary transformation $\ket{\Psi} \to \exp \sqof{ i \of{ \Delta / 2} \sum_j \sigma_{jS}^z t } \ket{\Psi}$. Our rotating frame Hamiltonian $H'$ is then given by
\begin{eqnarray}
H' &=&  \sum_j \sqof{-\frac{\omega}{2} \of{ \sigma_{j}^{z} + \sigma_{jS}^z }+ g \of{\sigma_{j}^+ \sigma_{jS}^- + \sigma_{j}^- \sigma_{jS}^+} }.
\end{eqnarray}

For the moment we consider $g = 0$. Modeling the environment as a thermally populated bath of harmonic oscillators with energies near $\omega$ and $\omega_S$ \footnote{C. Gardiner and P. Zoller, \emph{Quantum Noise: A Handbook of Markovian and Non-Markovian Quantum Stochastic Methods with Applications to Quantum Optics} (Springer, 2004)}, we note that at thermal equilibrium, the density of excited qubits $\rho^+$ is proportional to the ratio of the excitation and relaxation rates for the qubits to exchange excitations with the bath, so that
\begin{eqnarray}
\rho^+ = \frac{\Gamma_P^+}{\Gamma_P^-},
\end{eqnarray}
Here, $\Gamma_P^- = \of{ 1 + N_{th}} \kappa$ and $\Gamma_P^+ = N_{th} \kappa$, where $\kappa$ is the qubit-bath coupling and $N_{th} = e^{- \hbar \omega / k_B T}$ is the (appropriately normalized) density of excitations at the relevant energy in the thermal bath, which is exponentially small at low temperatures. 

On the other hand, the shadow qubit excitation density $\rho_{S}^+$ will be proportional to $e^{- \hbar \omega_S / k_B T}$, despite that the two groups of qubits are isoenergetic in (\ref{Hfreq}). This is because, as described in \cite{kapitchalker2014} and elsewhere, when considering continuously driven, open quantum systems, the bath degrees of freedom must be transformed in the same manner as the qubits. In this frame, both sets of bath modes now have frequency $\omega$, and the qubit-bath couplings $\of{ \sigma_{j(S)}^+ b_{(S)} + \sigma_{j(S)}^- b_{(S)}^\dagger }$ will be unchanged, though any \textit{static} couplings between qubits and bath modes at different energies will be rapidly oscillating and can thus be neglected. This is of course natural because moving to the rotating frame is just a unitary transformation used to simplify the analysis of the system's dynamics, but the key point is that if $\Delta \gg k_B T$, the shadow qubit population will be exponentially smaller and they can in turn act as a zero temperature bath for the primary qubits. A quick Fermi's golden rule calculation of the two-step process where an excitation in the primary qubits tunnels to a shadow qubit and decays into the oscillator bath yields an additional contribution to the primary qubit loss rate $\Gamma_P^-$: 
\begin{eqnarray}
\Gamma_P^- \to \Gamma_P^- + \frac{4 g^{2} \Gamma_S}{4 g^{2} + \Gamma_{S}^2}.
\end{eqnarray}
As the equilibrium population is still given by the ratio of the excitation and relaxation rates, the new effective system temperature is given by
\begin{eqnarray}
e^{- \hbar \omega / k_{B} T_{\rm eff} } = \kappa e^{- \hbar \omega / k_B T} / \of{\kappa + \frac{4 g^{2} \Gamma_S}{4 g^{2} + \Gamma_{S}^2} }.
\end{eqnarray}
In realistic devices the induced decay rate can be two to three orders of magnitude larger than the system-bath coupling $\kappa$, so while the reduction in effective temperature is only logarithmic, it can still be substantial. One obvious use for this mechanism is to reduce the error rate in preparing ancilla qubits-- most quantum algorithms require many auxiliary qubits to be introduced at various stages in the algorithm, always in a known state, so a passive cooling scheme could dramatically reduce the chance that an ancilla qubit is initialized in state $\ket{1}$ rather than $\ket{0}$ as intended, without having to perform the extra step of measuring the qubit directly. Of course, the passive cooling mechanism must be turned off while the qubit is being used for computation, but as the coupling is entirely generated by a drive field this is easy to do.

Provided that the shadow qubit detunings can be appropriately controlled, this system can also be used to cool strongly interacting many-qubit arrays, with the shadow qubit energies tuned to match the energy cost of excitations about the ground state. However, we caution that for generic models and more general couplings than simple frequency conversion, the shadow qubits will induce a finite error rate of their own which is only quadratically suppressed in the ratio of the couplings to the energy cost, $g/\omega$. To see this, we consider the previous example, but replace the exchange coupling by a full $xx$ term:
\begin{eqnarray}
H'' &=&  \sum_j \sqof{-\frac{\omega}{2} \of{ \sigma_{j}^{z} + \sigma_{jS}^z }+ g \sigma_{j}^x \sigma_{jS}^x }.
\end{eqnarray}
The $g$ term now introduces an error channel, in which the $xx$ spontaneously excites both qubits and the shadow qubit decays, leaving the primary qubit excited. Including this term, the effective temperature now reads:
\begin{eqnarray}
e^{- \hbar \omega / k_{B} T_{\rm eff} } \simeq \frac{ \kappa e^{- \hbar \omega / k_B T} + \frac{g^{2} \Gamma_S}{4 \omega^{2}}    }{\kappa + \frac{4 g^{2} \Gamma_S}{4 g^{2} + \Gamma_{S}^2} }.
\end{eqnarray}
For low enough temperatures (or large enough $g/\omega$) the induced error rate can be larger than the exponentially suppressed coupling to the outside world, and the shadow lattice can actually heat the primary qubits. One must therefore be careful when designing passive cooling systems to appropriately suppress the induced error rate, either by ensuring that the coupling Hamiltonian can only correct errors (as in the simple case of independent spins) or by making the coupling weak enough that the induced error rate is minimal. Whether or not passive cooling below ambient temperature can be implemented successfully thus depends on the detailed properties of the model and ratio of its excitation gap to the environment temperature.

\section{Higher-order interactions without gadgets}

As a second application of these couplings, we will now demonstrate how one can map superconducting devices to linear, $n$-level systems. For concreteness, we will consider the transmon qubit pair in FIG.~1(a); the two qubits have energy levels $\cuof{0,\omega_{i}, 2 \omega_{i} - \alpha_{i},3 \omega_{i} - \beta_{i}...}$. We further define:
\begin{eqnarray}
\Delta \equiv \omega_2 - \omega_1 ; \; \alpha_{ij} &\equiv & \alpha_i - \alpha_j ; \; P_{i}^{n} \equiv \ket{n_{i}} \bra{n_{i}}.
\end{eqnarray} 
For simplicity, we will consider the case where we wish to simulate bosons hopping between a pair of sites with local three-body interactions:
\begin{eqnarray}\label{H3B}
H &=& \omega_1 \of{a_1^\dagger a_1 + a_2^\dagger a_2} - g \of{a_1^\dagger a_2 + a_2^\dagger a_1} \\
& & + \frac{1}{6} \sqof{ U_1 \of{a_{1}^\dagger}^{3}\of{a_{1}}^{3} + U_2  \of{a_{2}^\dagger}^{3}\of{a_{2}}^{3}  }. \nonumber
\end{eqnarray}
By default, there is a strong 2-body interaction (coming from the $\alpha_{i}$ nonlinearity) which we must eliminate, and the qubits are detuned from each other so no tunneling occurs without an applied drive signal. To enable photon tunneling and eliminate the 2-body interaction, we introduce the following drive signal for the flux bias $F \of{t}$ in (\ref{bareint}):
\begin{eqnarray}\label{3bodyflux}
F \of{t} &=& \frac{\pi}{2} + k_0 \cos \of{\Delta t} \\
& & + k_1 \cos \of{ \of{\Delta + \alpha_1} t} + k_2 \cos \of{ \of{\Delta - \alpha_2} t} \nonumber \\
& & + k_3 \sqof{ \cos \of{ \of{\Delta + \alpha_{12}} t} +  \cos \of{ \of{\Delta - \alpha_{12}} t} }; \nonumber \end{eqnarray}
We now move to a rotating frame via the following transformation:
\begin{eqnarray}
\ket{\Psi} \to \exp \sqof{i \of{-\alpha_{1} P_{1}^{2} + \Delta a_{2}^\dagger a_{2} - \alpha_{2} P_{2}^{2}}t } \ket{\Psi}.
\end{eqnarray}
In this frame, the diagonal part of $H$ is simply $\omega_{1} \of{a_{1}^\dagger a_1 + a_{2}^\dagger a_2 } + U_1 P_{1}^{3} + U_2 P_{2}^{3}$, where $U_i = - \beta_i$; the two-photon nonlinearities have been eliminated in the unitary transformation. If, as before, we identify $\sin \phi_i = s \of{a_{i}^\dagger + a_{i}}$, the tunneling terms will be time-dependent, but for each tunneling term, there exists a component of (\ref{3bodyflux}) which eliminates the time dependence. For example, the $k_0 \cos \Delta t$ term removes the time dependence from $\ket{01}\bra{10} + \ket{10}\bra{01}$, $k_1 \cos \of{ \of{\Delta + \alpha_1} t} + k_2 \cos \of{ \of{\Delta - \alpha_2} t}$ removes the time dependence from $\ket{02}\bra{11} + \ket{11}\bra{02}$ and $\ket{20}\bra{11} + \ket{11}\bra{20}$, and so on. Provided that the energy spacings $\alpha_i$ and $\Delta$ are all large compared to the net tunneling matrix element $g$, the rapidly oscillating terms can be ignored and we obtain the desired Hamiltonian (\ref{H3B}). Further, we have achieved this without increasing the complexity of the physical circuit (as in perturbative gadget constructions such as \cite{kapitsimon}).

As in all previous examples, the phases of the signals at different frequencies must remain locked to avoid generating unwanted complex phases in the rotating frame tunnel couplings. Higher order interactions can be generated by adding additional signal components to cancel out the higher nonlinearities (such as the 3-body nonlinearity $\beta_i$). Likewise, the natural 2-body interaction can be reintroduced, with freely tunable magnitude and sign, by simply detuning the drive signals away from the $\alpha_i$. Note also that when counter-rotating terms and other higher order processes are taken into account, residual nonlinearities will likely remain, though for realistic parameters they will be 2-3 orders of magnitude smaller than the bare nonlinearity of the device.

\section{Conclusion}

We have demonstrated that a flux-biased Josephson junction can act as a universal interaction resource in the driven arrays of superconducting qubits, allowing arbitrary spin models to simulated dynamically by simply tuning applied drive fields (and leaving the physical wiring unchanged). We have also shown how most of this flexibility can be replicated through simple capacitive couplings between qubits with internal flux lines, potentially simplifying the construction, and demonstrated that the frequency-converting nature of the couplings can be used to passively cool qubits below the temperature of their surroundings. Finally, we have shown how incorporating multiple frequency components in a frequency-converting tunnel coupling can selectively eliminate qubit nonlinearities in the rotating frame, and thus generate higher-order local interactions than simple 2-body terms. Our results thus considerably expand the ``toolbox" for quantum simulation in superconducting qubit arrays, and demonstrate that extremely diverse classes of interacting spin models could be simulated with the same physical hardware.

\section{Acknowledgements}
I would like to thank M. Hafezi, J. Koch, Y. Lu, V. Oganesyan, D. Schuster and S. H. Simon for useful discussions. This material is based on work which was supported by the Graduate Center of the City University of New York, EPSRC Grant Nos. EP/I032487/1 and EP/I031014/1, and Worcester College of the University of Oxford.

\bibliography{biblio,EC_bib,SLbib}

\end{document}